\documentclass[aps,prl,twocolumn,amsmath,amssymb,floatfix]{revtex4-2}

\usepackage[utf8]{inputenc}
\usepackage{amsmath,amssymb,amsfonts}
\usepackage{bm}
\usepackage{physics}
\usepackage{tikz}
\usepackage{microtype}
\usepackage[colorlinks=true, allcolors=blue]{hyperref}
\usepackage{tikz}
\usepackage{pgfplots}
\pgfplotsset{compat=1.18}

\begin{document}
	
	\title{Classical Coherence and Biological Aging}
	\author{Yehuda Roth}
	\affiliation{Faculty of Sciences, Oranim College, Israel}
	\date{\today}
	
	\begin{abstract}
		In previous work it was argued that the cells of a multicellular organism
		form a classically coherent system and that such coherence is essential
		for life. Here this claim is made precise by introducing an explicit
		classical formalism in which a driven, dissipative many-cell system is
		represented by a single state vector in an abstract code space. Using
		Dirac's bra--ket notation purely as a compact representation of
		classical states, we construct an analogue of the center-of-mass
		coordinate that encodes the organismal identity and show how a common
		code shared by all cells corresponds to a coherent, low-entropy phase in
		this space. We then map this structure onto DNA sequence space by
		introducing a classical ``Biological Hamiltonian'' whose generalized
		coordinates encode genetic codes and their cell-wise distribution, so
		that the organismal identity is represented by a global code state
		rather than by individual molecular constituents. Within this framework
		we define a time-dependent linear operator with code-correcting and
		code-breaking terms, weighted by coefficients $A(t)$ and $B(t)$, which
		captures the balance between restorative dynamics and
		environment-induced damage to the code. Aging is described as a slow
		drift in these control parameters: as $A(t)$ decreases and $B(t)$
		increases, the coherent identity state becomes progressively less
		stable, and the organism moves from a regime of robust code coherence to
		a regime dominated by stochastic code variability. In this picture,
		death appears as a dissipative transition in which the global identity
		state can no longer be maintained, so that the organismal coherence
		collapses despite ongoing material turnover. The resulting framework
		suggests experimentally testable signatures in aging multicellular
		systems, such as a late-life loss of restoration capacity and an abrupt
		breakdown of code coherence at the end-of-life transition.
	\end{abstract}
	
	\maketitle

	\section{Introduction}
	Coherence is a central property of quantum theory responsible for
	fascinating inversions such as the entanglement property.\cite{Marais2018}
	For example, in EPR pairs, two spin-$\tfrac{1}{2}$ particles, even when very far
	apart, lose their independence and are defined as a single entity with
	total spin zero. In other words, coherence in physics is a collective
	property, causing the participating objects to belong to a common entity
	by losing their individual identity.
	
	A multicellular organism is just such a system: its constituent cells
	lose their effective independence and belong to the identity of the
	organism. This makes it natural to treat the organism itself as a
	coherent entity---its physical identity.\cite{Davies2021,Roth2026ClassicalCoherence}
	We present here a hypothesis that a living organism preserves coherence,
	while loss of coherence means death, in line with recent attempts to
	distinguish organisms from colonies by classical coherence
	criteria.\cite{Roth2026ClassicalCoherence}
	
	There are only a few cases in the professional literature of classical
	physics describing coherence between macroscopic objects, so that
	phenomena exhibiting collective behavior are often almost automatically
	framed as ``quantum'' and discussed in terms of quantum
	coherence.\cite{Haken1983,anderson1972} In this context, for example,
	Fr{\"o}hlich suggested that driven biological structures such as cell
	membranes and polar biomolecular assemblies might support long-range
	coherent electrical vibrations, somewhat analogous to Bose--Einstein
	condensations of vibrational states, and thus invited a quantum-like
	interpretation of biological order.\cite{Frohlich1968} However, applying
	such a quantum description to a warm, wet biological environment is
	problematic: at physiological temperatures, strong coupling to the
	surrounding thermal bath leads to rapid decoherence and tends to wash
	out the fragile quantum phase relations, so that any true quantum
	coherence would have to survive on time scales and length scales that
	are difficult to accommodate within the noisy, strongly dissipative
	conditions inside living cells.\cite{Frohlich1968,Marais2018}
	
	In order to build a description of coherence in non-quantum systems that
	will not suffer from coherence-destroying phenomena, we propose the
	classical center-of-mass system as a model for defining coherence
	between objects (particles), as presented earlier.\cite{roth2019superposition}
	Although its mathematical formulation is not the same as the quantum
	one, it is still described by a superposition of position vectors and,
	within the framework of the definition of the system, the identity of a
	particle as an individual disappears.\cite{Roth2026ClassicalCoherence}
	First, we adapt the classical formalism of the center-of-mass system to
	quantum notation. Then we show an analogy between the center-of-mass
	system and the DNA codes in a multicellular organism, thereby presenting
	classical coherence between the DNA codes of a multicellular
	organism.\cite{Roth2026ClassicalCoherence,Goldschmidt2014}
	
	Finally, using an operator formalism adapted to our classical
	description of the DNA code, we present the influence of the
	environment through a coherence-breaking operator and an error-correcting
	operator acting on the code.\cite{Roth2026ClassicalCoherence}
	\section{Classical coherence presented in Dirac Notation}
	\label{sec:classical_coherence_fock}
	We present a classical formalism for describing coherence. However,
	since coherence between objects is discussed mainly in quantum theory,
	the common formalism for such descriptions is Dirac notation. We
	therefore present our classical system by means of Dirac's bra--ket
	notation.
	
	Consider a system of $N$ particles labeled by $i$, where each particle
	$i$ has an energy $E_i$. We can then associate each particle with a
	state $\{\lvert i \rangle\}$, which not only labels the particle but also
	satisfies the orthogonality relation $\langle i \vert j \rangle=\delta_{i,j}$.
	This definition of the states has nothing to do with quantum mechanics
	and is purely formal.
	
	If a particle $i$ possesses an energy $E_i$, we can express the classical
	Hamiltonian of the multi-particle system in spectral form:
	\begin{equation}
		H = \sum_i E_i \, \lvert i \rangle \langle i \rvert,
		\label{eq:H_spectral_classical}
	\end{equation}
	This expression is formally identical to the spectral
	decomposition of a quantum Hamiltonian, but here the $\lvert i \rangle$
	denote a formal \textit{classical} basis states rather than quantum eigenstates. 
		
	In classical mechanics, the transition to the center-of-mass coordinate negates the individuality of the particles as they are all represented by a single coordinate - the mass-center coordinate. Here, we show how to define the center-of-mass system by the states of the individual particles $\left\{\lvert i\rangle\right\}$.\\
	We start with the following state,
	\begin{equation}
		\lvert q \rangle = \sum_{i=1}^N q_i \lvert i \rangle,
		\label{eq:config_vector}
	\end{equation}
	which encodes the generalized coordinates $q_i$ of all constituents in a
	single abstract state. In this picture, the conventional view in which
	each particle $i$ carries its own distinguishable coordinate $q_i$ is
	replaced by a global coordinate vector $\lvert q \rangle$ that associates
	the system as a whole with its generalized configuration. Individual
	coordinates $q_i$ are meaningful only as components of this global state.
	
	We also define the mass-weight vector
	\begin{equation}
		\lvert w \rangle
		= \sum_{i=1}^N w_i \lvert i \rangle,
		\qquad
		w_i = \frac{m_i}{M},
		\qquad
		M = \sum_{i=1}^N m_i,
		\label{eq:weight_vector}
	\end{equation}
	which represents the relative impact of each particle in the collective
	description. In system language, one might say that “the contribution of
	particle $i$ is $m_i/M$ of the total,” so that an individual mass outside
	this context becomes meaningless. We emphasize that both vectors
	$\lvert q \rangle$ and $\lvert w \rangle$ \textbf{are not normalized}.
	
	In this notation, the standard center-of-mass coordinate
	is precisely the projection of the mass-weight vector onto the
	configuration vector,
	\begin{equation}
		Q_q
		= \langle w \vert q \rangle
		= \sum_{i=1}^N w_i^\ast q_i
		= \sum_{i=1}^N \frac{m_i}{M}\, q_i.
		\label{eq:com_projection}
	\end{equation}
	Thus, the usual center-of-mass position is recovered as the scalar
	component of $\lvert q \rangle$ along the mass-weight direction
	$\lvert w \rangle$ in particle-label space.
	
	For identical particle locations, $q_i = q_0$ for all $i$, substituting
	into Eq.~\eqref{eq:com_projection} yields
	\begin{equation}
		Q_{q_0} = q_0,
		\label{eq:com_identical}
	\end{equation}
	In other words, the common
	coordinate coincides with the center-of-mass coordinate. 
		
	At first sight
	this might suggest that no coherence is present in such a scenario.
	However, even when all particles share the same coordinate, the weight
	vector of Eq.~\eqref{eq:weight_vector} remains well defined and encodes
	the coherent organization of the system. Coherence also remains through Eq.~\ref{eq:config_vector} that now becomes
	\begin{equation}
		\lvert q \rangle = q_0\sum_{i=1}^N  \lvert i \rangle,
		\label{eq:config_vecto1r}
	\end{equation}
	
	At first sight, a configuration in which all particles occupy the same
	point in configuration space may seem artificial from the perspective
	of ordinary mechanics. In our biological model, introduced in
	Sec.~\ref{sec:biological_model}, the coordinates $q_i$ are not literal
	spatial positions but abstract labels of DNA codes, while the cells
	carrying these codes are located at different spatial points. Thus, a
	state with $q_1 = \dots = q_N$ should be understood as a situation in
	which all cells share the same code, not as a collapse of all particles
	to the same physical location. In this sense, the center-of-mass
	construction above captures coherence in the space of codes, while the
	underlying spatial separation of the cells remains implicit in the
	classical description.
	
	As mentioned, the center-of-mass state does not belong to a spanning set
	of a Hilbert space, and we are therefore not obliged to define $N-1$
	additional states that complete such a basis. Nevertheless, in ordinary
	classical mechanics complementary coordinates are associated with
	relative coordinates. In our framework, all particles are located at the
	same position in code space, so there is no need to introduce relative
	coordinates explicitly.
	
	Our framework is entirely classical. In analogy with classical
	mechanics, we assign effective energy values to DNA codes and then
	apply the same formalism used for the transition to the center-of-mass
	coordinate, where in the biological context the collective coordinate
	is the genetic code. Once the biological energy description is
	accepted, the subsequent construction is purely technical, which
	supports confidence in the correctness of the model. The use of state
	vectors and operators in code space is simply a convenient way to
	express this classical description and does not rely on any quantum
	effects.
\section{The biological coherence model}
\label{sec:biological_model}

We introduce the Hamiltonian of an $N$-cell DNA-coded system as
\begin{equation}
	H_{\mathrm{code}}
	= \sum_{i=1}^N E_{g_i} \, \lvert i \rangle \langle i \rvert,
	\label{eq:H_code_spectral}
\end{equation}
where $g_i$ is a variable that represents the genetic code of the
$i$-th cell. We associate to each DNA code $g_i$ an effective energy
$E_{g_i}$, which plays the role of a free-energy level in code space
and quantifies the relative stability of that code. The biological
coherence model remains entirely classical, as it is built directly on
energy states and transition rates, and relies only on a simple analogy
to classical mechanics to describe the organism as a many-body system
in code space.

Equations~\eqref{eq:H_spectral_classical} and \eqref{eq:H_code_spectral}
share the same structural form: in both cases a many-body configuration
is represented as a vector in an abstract label space. In direct analogy
with the center-of-mass formalism, the common location in coordinate
space corresponds here to an identical code in code space, which allows
us to introduce a notion of classical coherence at the organismal level.

We define the code vector
\begin{equation}
	\lvert g \rangle = \sum_{i=1}^N g_i \lvert i \rangle,
	\label{eq:code_vector}
\end{equation}
which can be associated with the identity of the system. In addition,
we define a distribution vector
\begin{equation}
	\lvert P \rangle = \sum_{i=1}^N P_{g_i} \lvert i \rangle,
	\label{eq:distribution_vector1}
\end{equation}
where $P_{g_i}$ is the probability of finding the DNA of cell $i$
in the code labeled by $g_i$. For a set of $K$ possible codes,
the number of distinct assignment configurations $\lvert P \rangle$
for $N$ cells is $K^N$. Note that these configurations are classical
distribution states in code space and, in general, are not mutually
orthogonal.

In the ideal coherent limit, all cells are aligned with the same code
and the distribution vector $\lvert P \rangle$ is uniform across cells.
Deviations from this homogeneous situation, such as cells that have
drifted away from the dominant code, are encoded in the non-uniform
components $P_{g_i}$. In this way, the vector $\lvert P \rangle$
captures the degree to which the organism departs from a perfectly
coherent code configuration.

By direct analogy with the center-of-mass coordinate, we define the code
vector $g$ as
\begin{equation}
	\langle g \rangle
	= \langle P \vert g \rangle
	= \sum_{i=1}^N P_{g_i} \, g_i,
	\label{eq:com_projection_code}
\end{equation}
which is the average code value of the organism weighted by the cell-wise
probabilities.

For the ideal scenario of a single code $g_0$ we obtain
\begin{equation}
	\lvert g_0 \rangle = g_0 \sum_{i=1}^N \lvert i \rangle,
	\label{eq:code_vector_single}
\end{equation}
corresponding to the situation in which all cells share the same code.
In this case the distribution vector becomes
\begin{equation}
	\lvert P_0 \rangle = \sum_{i=1}^N P_{g_0} \lvert i \rangle,
	\label{eq:distribution_vector_single}
\end{equation}
where $P_{g_0}$ is the probability assigned to the unique code
$g_0$. If we assume that every cell certainly carries the code
$g_0$, we have $P_{g_0}=1$ and therefore
\begin{equation}
	\lvert P_0 \rangle = \sum_{i=1}^N \lvert i \rangle,
	\label{eq:P_single_uniform}
\end{equation}
so that the expectation value reduces to
\begin{equation}
	\langle g \rangle
	= \langle P \vert g \rangle
	= g_0 \sum_{i=1}^N 1
	= N \, g_0.
	\label{eq:expectation_single_raw}
\end{equation}
The vector $\lvert P_0 \rangle$ of Eq.~\eqref{eq:P_single_uniform}
describes a living configuration in which all cells share the same
code, independently of the particular organism identity (i.e. without
reference to the specific value of $g_0$). By contrast, $\lvert g_0
\rangle$ of Eq.~\eqref{eq:code_vector_single} specifies that common code
value and thus encodes the organism’s identity.

In realistic biological settings, material turnover, replication errors
and environmental damage continuously perturb the distribution vector
away from the perfectly coherent configuration. Aging and organismal
decline can therefore be viewed, within this classical framework, as
processes in which the system gradually loses coherence in code space,
until no global state $\lvert g_0 \rangle$ can be robustly maintained.
\section{Cells Aging and Death}
During the lifetime of a cell, codes are continuously both damaged and
repaired. To capture how a multicellular organism actively maintains its coherent
identity state against continuous code damage, we introduce a classical
maintenance operator $\mathbb{O}$ acting on the code space.
\begin{equation}
	\mathbb{O}
	= A(t)\,\!\!\!\!\!\!\underbrace{\lvert g_0\rangle\langle g\rvert}_{\begin{array}{c}
			\text{code}\\[-5pt]
			\text{correcting}\\[-5pt]
			\text{term}
	\end{array}}\,\!\!\!\!\!\!
	+ B(t)\,\,\!\!\!\!\!\!\underbrace{\lvert g\rangle\langle g_0\rvert}_{\begin{array}{c}
			\text{code}\\[-5pt]
			\text{breaking}\\[-5pt]
			\text{term}
	\end{array}}.
\label{mathO}
\end{equation}
Here $\lvert g\rangle$ denotes the current (possibly perturbed) code state in
the abstract code space, while $\lvert g_0\rangle$ represents the ideal
reference code associated with the organismal identity. The operator
$\mathbb{O}$ acts on this code space as a purely classical linear map:
the first term, weighted by $A(t)$, pulls the system back toward the
reference code and thus models restorative, error-correcting processes;
the second term, weighted by $B(t)$, pushes the reference code toward
perturbed configurations and thus models the code-breaking influence of
the environment. The time-dependent coefficients $A(t)$ and $B(t)$
therefore quantify, at each moment, the relative strength of
code-correcting versus code-breaking dynamics for a given cell.

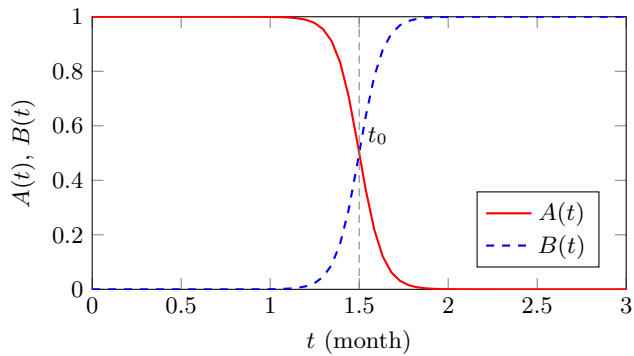
\begin{figure}[t]
		\centering
		\begin{tikzpicture}
			\begin{axis}[
				width=\columnwidth,
				height=0.6\columnwidth,
				xmin=0, xmax=3,
				ymin=0, ymax=1,
				xlabel={$t$ (month)},
				ylabel={$A(t),\,B(t)$},
				legend style={at={(0.95,0.08)},anchor=south east},
				domain=0:120,
				samples=2500,
				]
				
				\pgfmathsetmacro{\tzero}{1.5}   
				\pgfmathsetmacro{\kappa}{15} 
				
				\addplot[red,thick]
				{1/(1 + exp(\kappa*(x - \tzero)))};
				\addlegendentry{$A(t)$}
				
				\addplot[blue,thick,dashed]
				{1 - 1/(1 + exp(\kappa*(x - \tzero)))};
				\addlegendentry{$B(t)$}
				
				\addplot[gray,densely dashed]
				coordinates {(\tzero,0) (\tzero,1)};
				\node[black,anchor=south west] at (axis cs:\tzero,0.5)
				{$t_0$};
				
			\end{axis}
		\end{tikzpicture}
		\caption{Schematic time dependence of the weights $A(t)$ and $B(t)$ of a cell
			over a 3-year lifespan. The curves cross at the effective
			lifetime $t_0$, where $A(t_0)=B(t_0)=1/2$. Aging begins around
			midlife (e.g.\ $t\approx 1.2$ months).}
		\label{fig:AB_lifetime}
	\end{figure}

In this work we do not derive explicit microscopic expressions for
$A(t)$ and $B(t)$; instead, we adopt a simple phenomenological choice in
which their time dependence follows complementary sigmoidal curves, as
shown schematically in Fig.~\ref{fig:AB_lifetime}. This captures the
idea that repair capacity (encoded in $A(t)$) gradually declines with
age, while damage-accumulating influences (encoded in $B(t)$) become
increasingly dominant. Early in the cell's life $A(t)$ dominates over
$B(t)$, so the organism remains in a low-entropy coherent phase,
whereas an increase of $B(t)$ relative to $A(t)$ drives the system
toward decoherence of the DNA code and, eventually, loss of organismal
identity. Just as the standard expression for the energy of a system
provides a symbolic description without specifying the detailed
mechanisms by which that energy is generated, our operator model does
not attempt to resolve the microscopic pathways of code damage and
repair, but only captures them in a compact mathematical form.
	
	\section{Discussion and outlook}
	
	In a previous work, we presented the idea that the DNA codes in a
	multicellular organism are coherent, in the sense that they do
	not possess independence beyond belonging to the organism as a whole,
	and we proposed that the loss of this coherence entails the death of the
	organism.\cite{Roth2026ClassicalCoherence} In the present paper we have
	given this idea a precise classical formulation. Using Dirac's bra--ket
	notation purely as a compact representation of classical states, we
	constructed an analogue of the center-of-mass coordinate that encodes
	organismal identity and showed that a common code shared by all cells
	corresponds to a coherent state.
	
	On this basis we introduced a classical ``Biological Hamiltonian'' whose
	generalized coordinates encode the DNA codes, so that the organismal
	identity is represented by a global code state rather than by individual
	molecular constituents. Within this framework we defined a time-dependent
	operator with code-correcting and code-breaking terms, weighted by
	coefficients $A(t)$ and $B(t)$, which captures the balance between
	restorative dynamics and environment-induced damage to the code. Aging is described as a slow drift in these control parameters: as
	$A(t)$ decreases and $B(t)$ increases, the balance shifts from
	code-correcting to code-breaking dynamics, and the cell’s code state
	becomes progressively less faithful to its reference code $g_0$.\cite{Roth2026ClassicalCoherence}
	At first glance, the operator defined in Eq.~\ref{mathO} by inspecting a
	single cell seems to have nothing to do with coherence. However, since
	it is constructed from the code vector $\lvert g \rangle$ of
	Eq.~\ref{eq:code_vector}, which is a superposition of the cells' codes,
	we can say that $\mathbb{O}$ acts within the coherence principle: it
	modifies the organismal code state by applying code-correcting and
	code-breaking processes to a collectively defined superposition of
	cell-specific codes.
	
	In this picture, identity is tied to functional occupancy rather than to
	specific molecules and persists as long as restoration dynamics keep the
	system close to its coherent code state despite continuous material
	turnover.\cite{Maturana1980,Thompson2007} Organismal coherence is
	therefore a property of the dynamically maintained pattern in sequence
	and state space, not of any particular set of constituent parts. This
	view aligns with classical notions of autopoiesis while embedding them
	in an explicit mechanical and information-theoretic framework for aging
	and death.\cite{Maturana1980,Thompson2007} It also extends previous work
	on classical superposition and coherence in single systems\cite{roth2019superposition}
	to multicellular active matter, where the key degrees of freedom are
	genetic and functional rather than purely mechanical.
	
	Beyond its conceptual implications, viewing a multicellular organism as
	a classically coherent code state has concrete scientific and
	technological advantages. Scientifically, the framework makes it possible
	to pose sharp questions about when an organism retains or loses its
	identity in terms of trajectories in code space, rather than in terms of
	particular molecules or isolated pathways, and connects classical coherence
	criteria for organisms to quantitative measures of stability in sequence
	space.\cite{Roth2026ClassicalCoherence,roth2019superposition} Technologically,
	the model suggests new types of observable quantities: for example,
	coarse-grained indices of code coherence that track how close the
	organismal state $\lvert g\rangle$ remains to a reference identity state
	$\lvert g_0\rangle$, or effective proxies for the balance between restoration
	and damage encoded in the coefficients $A(t)$ and $B(t)$.\cite{Roth2026ClassicalCoherence}
	Such indicators could, in principle, be extracted from high-throughput
	sequencing and single-cell data and used to identify late-life regimes in
	which restoration capacity collapses and code variability accelerates,
	providing candidate biomarkers for aging and for the onset of degenerative
	failure in multicellular systems.\cite{Davies2021,Sender2016Turnover} More
	broadly, the framework suggests that classical coherence in active
	multicellular matter may offer a useful language for describing diverse
	biological phenomena, from morphogenesis and regeneration to aging and
	degenerative failure, by treating the maintenance and loss of organismal
	identity as instances of a single dissipative transition in code space.
	%
	

\begin{thebibliography}{14}%
		\makeatletter
		\providecommand \@ifxundefined [1]{%
			\@ifx{#1\undefined}
		}%
		\providecommand \@ifnum [1]{%
			\ifnum #1\expandafter \@firstoftwo
			\else \expandafter \@secondoftwo
			\fi
		}%
		\providecommand \@ifx [1]{%
			\ifx #1\expandafter \@firstoftwo
			\else \expandafter \@secondoftwo
			\fi
		}%
		\providecommand \natexlab [1]{#1}%
		\providecommand \enquote  [1]{``#1''}%
		\providecommand \bibnamefont  [1]{#1}%
		\providecommand \bibfnamefont [1]{#1}%
		\providecommand \citenamefont [1]{#1}%
		\providecommand \href@noop [0]{\@secondoftwo}%
		\providecommand \href [0]{\begingroup \@sanitize@url \@href}%
		\providecommand \@href[1]{\@@startlink{#1}\@@href}%
		\providecommand \@@href[1]{\endgroup#1\@@endlink}%
		\providecommand \@sanitize@url [0]{\catcode `\\12\catcode `\$12\catcode
			`\&12\catcode `\#12\catcode `\^12\catcode `\_12\catcode `\%12\relax}%
		\providecommand \@@startlink[1]{}%
		\providecommand \@@endlink[0]{}%
		\providecommand \url  [0]{\begingroup\@sanitize@url \@url }%
		\providecommand \@url [1]{\endgroup\@href {#1}{\urlprefix }}%
		\providecommand \urlprefix  [0]{URL }%
		\providecommand \Eprint [0]{\href }%
		\providecommand \doibase [0]{https://doi.org/}%
		\providecommand \selectlanguage [0]{\@gobble}%
		\providecommand \bibinfo  [0]{\@secondoftwo}%
		\providecommand \bibfield  [0]{\@secondoftwo}%
		\providecommand \translation [1]{[#1]}%
		\providecommand \BibitemOpen [0]{}%
		\providecommand \bibitemStop [0]{}%
		\providecommand \bibitemNoStop [0]{.\EOS\space}%
		\providecommand \EOS [0]{\spacefactor3000\relax}%
		\providecommand \BibitemShut  [1]{\csname bibitem#1\endcsname}%
		\let\auto@bib@innerbib\@empty
		\bibitem [{\citenamefont {Marais}\ \emph {et~al.}(2018)\citenamefont {Marais}
			\emph {et~al.}}]{Marais2018}%
		\BibitemOpen
		\bibfield  {author} {\bibinfo {author} {\bibfnamefont {A.}~\bibnamefont
				{Marais}} \emph {et~al.},\ }\href {https://doi.org/10.1098/rsif.2018.0640}
		{\bibfield  {journal} {\bibinfo  {journal} {J. R. Soc. Interface}\ }\textbf
			{\bibinfo {volume} {15}},\ \bibinfo {pages} {20180640} (\bibinfo {year}
			{2018})}\BibitemShut {NoStop}%
		\bibitem [{\citenamefont {Davies}\ \emph {et~al.}(2021)\citenamefont {Davies},
			\citenamefont {Walker},\ and\ \citenamefont {Kempes}}]{Davies2021}%
		\BibitemOpen
		\bibfield  {author} {\bibinfo {author} {\bibfnamefont {P.~C.~W.}\
				\bibnamefont {Davies}}, \bibinfo {author} {\bibfnamefont {S.~I.}\
				\bibnamefont {Walker}},\ and\ \bibinfo {author} {\bibfnamefont {C.~P.}\
				\bibnamefont {Kempes}},\ }\href {https://doi.org/10.1038/s42254-021-00366-y}
		{\bibfield  {journal} {\bibinfo  {journal} {Nat. Rev. Phys.}\ }\textbf
			{\bibinfo {volume} {3}},\ \bibinfo {pages} {711} (\bibinfo {year}
			{2021})}\BibitemShut {NoStop}%
		\bibitem [{\citenamefont {Roth}(2026)}]{Roth2026ClassicalCoherence}%
		\BibitemOpen
		\bibfield  {author} {\bibinfo {author} {\bibfnamefont {Y.}~\bibnamefont
				{Roth}},\ }\bibfield  {journal} {\bibinfo  {journal} {arXiv preprint}\ }\href
		{https://doi.org/10.48550/arXiv.2606.02801} {10.48550/arXiv.2606.02801}
		(\bibinfo {year} {2026}),\ \Eprint {https://arxiv.org/abs/2606.02801}
		{arXiv:2606.02801 [physics.bio-ph]} \BibitemShut {NoStop}%
		\bibitem [{\citenamefont {Haken}(1983)}]{Haken1983}%
		\BibitemOpen
		\bibfield  {author} {\bibinfo {author} {\bibfnamefont {H.}~\bibnamefont
				{Haken}},\ }\href@noop {} {\emph {\bibinfo {title} {Synergetics: An
					Introduction}}},\ \bibinfo {edition} {3rd}\ ed.\ (\bibinfo  {publisher}
		{Springer-Verlag},\ \bibinfo {address} {Berlin},\ \bibinfo {year}
		{1983})\BibitemShut {NoStop}%
		\bibitem [{\citenamefont {Anderson}(1972)}]{anderson1972}%
		\BibitemOpen
		\bibfield  {author} {\bibinfo {author} {\bibfnamefont {P.~W.}\ \bibnamefont
				{Anderson}},\ }\href {https://doi.org/10.1126/science.177.4047.393}
		{\bibfield  {journal} {\bibinfo  {journal} {Science}\ }\textbf {\bibinfo
				{volume} {177}},\ \bibinfo {pages} {393} (\bibinfo {year}
			{1972})}\BibitemShut {NoStop}%
		\bibitem [{\citenamefont {Fr{\"o}hlich}(1968)}]{Frohlich1968}%
		\BibitemOpen
		\bibfield  {author} {\bibinfo {author} {\bibfnamefont {H.}~\bibnamefont
				{Fr{\"o}hlich}},\ }\href {https://doi.org/10.1002/qua.560020505} {\bibfield
			{journal} {\bibinfo  {journal} {Int. J. Quantum Chem.}\ }\textbf {\bibinfo
				{volume} {2}},\ \bibinfo {pages} {641} (\bibinfo {year} {1968})}\BibitemShut
		{NoStop}%
		\bibitem [{\citenamefont {Roth}(2019)}]{roth2019superposition}%
		\BibitemOpen
		\bibfield  {author} {\bibinfo {author} {\bibfnamefont {Y.}~\bibnamefont
				{Roth}},\ }\href {https://doi.org/10.1016/j.rinp.2019.102387} {\bibfield
			{journal} {\bibinfo  {journal} {Results Phys.}\ }\textbf {\bibinfo {volume}
				{14}},\ \bibinfo {pages} {102387} (\bibinfo {year} {2019})}\BibitemShut
		{NoStop}%
		\bibitem [{\citenamefont {Goldschmidt}(2014)}]{Goldschmidt2014}%
		\BibitemOpen
		\bibfield  {author} {\bibinfo {author} {\bibfnamefont {E.~E.}\ \bibnamefont
				{Goldschmidt}},\ }\href {https://doi.org/10.3389/fpls.2014.00273} {\bibfield
			{journal} {\bibinfo  {journal} {Front. Plant Sci.}\ }\textbf {\bibinfo
				{volume} {5}},\ \bibinfo {pages} {273} (\bibinfo {year} {2014})}\BibitemShut
		{NoStop}%
		\bibitem [{\citenamefont {Maturana}\ and\ \citenamefont
			{Varela}(1980)}]{Maturana1980}%
		\BibitemOpen
		\bibfield  {author} {\bibinfo {author} {\bibfnamefont {H.~R.}\ \bibnamefont
				{Maturana}}\ and\ \bibinfo {author} {\bibfnamefont {F.~J.}\ \bibnamefont
				{Varela}},\ }\href@noop {} {\emph {\bibinfo {title} {Autopoiesis and
					Cognition: The Realization of the Living}}},\ \bibinfo {series} {Boston
			Studies in the Philosophy of Science}, Vol.~\bibinfo {volume} {42}\ (\bibinfo
		{publisher} {D. Reidel},\ \bibinfo {address} {Dordrecht},\ \bibinfo {year}
		{1980})\BibitemShut {NoStop}%
		\bibitem [{\citenamefont {Thompson}(2007)}]{Thompson2007}%
		\BibitemOpen
		\bibfield  {author} {\bibinfo {author} {\bibfnamefont {E.}~\bibnamefont
				{Thompson}},\ }\href@noop {} {\emph {\bibinfo {title} {Mind in Life: Biology,
					Phenomenology, and the Sciences of Mind}}}\ (\bibinfo  {publisher} {Harvard
			University Press},\ \bibinfo {address} {Cambridge, MA},\ \bibinfo {year}
		{2007})\BibitemShut {NoStop}%
		\bibitem [{\citenamefont {Marchetti}\ \emph {et~al.}(2013)\citenamefont
			{Marchetti}, \citenamefont {Joanny}, \citenamefont {Ramaswamy}, \citenamefont
			{Liverpool}, \citenamefont {Prost}, \citenamefont {Rao},\ and\ \citenamefont
			{Simha}}]{Marchetti2013}%
		\BibitemOpen
		\bibfield  {author} {\bibinfo {author} {\bibfnamefont {M.~C.}\ \bibnamefont
				{Marchetti}}, \bibinfo {author} {\bibfnamefont {J.~F.}\ \bibnamefont
				{Joanny}}, \bibinfo {author} {\bibfnamefont {S.}~\bibnamefont {Ramaswamy}},
			\bibinfo {author} {\bibfnamefont {T.~B.}\ \bibnamefont {Liverpool}}, \bibinfo
			{author} {\bibfnamefont {J.}~\bibnamefont {Prost}}, \bibinfo {author}
			{\bibfnamefont {M.}~\bibnamefont {Rao}},\ and\ \bibinfo {author}
			{\bibfnamefont {R.~A.}\ \bibnamefont {Simha}},\ }\href
		{https://doi.org/10.1103/RevModPhys.85.1143} {\bibfield  {journal} {\bibinfo
				{journal} {Rev. Mod. Phys.}\ }\textbf {\bibinfo {volume} {85}},\ \bibinfo
			{pages} {1143} (\bibinfo {year} {2013})}\BibitemShut {NoStop}%
		\bibitem [{\citenamefont {Broedersz}\ and\ \citenamefont
			{MacKintosh}(2024)}]{Broedersz2024}%
		\BibitemOpen
		\bibfield  {author} {\bibinfo {author} {\bibfnamefont {C.~P.}\ \bibnamefont
				{Broedersz}}\ and\ \bibinfo {author} {\bibfnamefont {F.~C.}\ \bibnamefont
				{MacKintosh}},\ }\href {https://doi.org/10.1103/RevModPhys.96.015002}
		{\bibfield  {journal} {\bibinfo  {journal} {Rev. Mod. Phys.}\ }\textbf
			{\bibinfo {volume} {96}},\ \bibinfo {pages} {015002} (\bibinfo {year}
			{2024})}\BibitemShut {NoStop}%
		\bibitem [{\citenamefont {Fang}\ \emph {et~al.}(2019)\citenamefont {Fang},
			\citenamefont {Kruse}, \citenamefont {Lu},\ and\ \citenamefont
			{Wang}}]{Fang2020}%
		\BibitemOpen
		\bibfield  {author} {\bibinfo {author} {\bibfnamefont {X.}~\bibnamefont
				{Fang}}, \bibinfo {author} {\bibfnamefont {K.}~\bibnamefont {Kruse}},
			\bibinfo {author} {\bibfnamefont {T.}~\bibnamefont {Lu}},\ and\ \bibinfo
			{author} {\bibfnamefont {J.}~\bibnamefont {Wang}},\ }\href
		{https://doi.org/10.1103/RevModPhys.91.045004} {\bibfield  {journal}
			{\bibinfo  {journal} {Reviews of Modern Physics}\ }\textbf {\bibinfo {volume}
				{91}},\ \bibinfo {pages} {045004} (\bibinfo {year} {2019})}\BibitemShut
		{NoStop}%
		\bibitem [{\citenamefont {Sender}\ \emph {et~al.}(2016)\citenamefont {Sender},
			\citenamefont {Fuchs},\ and\ \citenamefont {Milo}}]{Sender2016Turnover}%
		\BibitemOpen
		\bibfield  {author} {\bibinfo {author} {\bibfnamefont {R.}~\bibnamefont
				{Sender}}, \bibinfo {author} {\bibfnamefont {S.}~\bibnamefont {Fuchs}},\ and\
			\bibinfo {author} {\bibfnamefont {R.}~\bibnamefont {Milo}},\ }\href
		{https://doi.org/10.1371/journal.pbio.1002533} {\bibfield  {journal}
			{\bibinfo  {journal} {PLoS Biology}\ }\textbf {\bibinfo {volume} {14}},\
			\bibinfo {pages} {e1002533} (\bibinfo {year} {2016})}\BibitemShut {NoStop}%
	\end{thebibliography}
	
\end{document}